\documentclass[runningheads]{llncs}
\usepackage[T1]{fontenc}
\usepackage{booktabs}
\usepackage{array}
\usepackage{graphicx}

\usepackage{color}
\begin{document}
%
\title{Security and Privacy in Virtual and Robotic Assistive Systems: A Comparative Framework}
%
%
\author{Nelly Elsayed\inst{1}\orcidID{0000-0003-0082-1450} 
}
\authorrunning{Elsayed et al.}
%
\institute{School of Information Technology, University of Cincinnati, OH, United States
}
\maketitle              
\begin{abstract}
Assistive technologies increasingly support independence, accessibility, and safety for older adults, people with disabilities, and individuals requiring continuous care. Two major categories are virtual assistive systems and robotic assistive systems operating in physical environments. Although both offer significant benefits, they introduce important security and privacy risks due to their reliance on artificial intelligence, network connectivity, and sensor-based perception. Virtual systems are primarily exposed to threats involving data privacy, unauthorized access, and adversarial voice manipulation. In contrast, robotic systems introduce additional cyber-physical risks such as sensor spoofing, perception manipulation, command injection, and physical safety hazards. In this paper, we present a comparative analysis of security and privacy challenges across these systems. We develop a unified comparative threat-modeling framework that enables structured analysis of attack surfaces, risk profiles, and safety implications across both systems. Moreover, we provide design recommendations for developing secure, privacy-preserving, and trustworthy assistive technologies.

\end{abstract}
%
%
%

\section{Introduction}

Assistive technologies have become an important component of modern intelligent systems, supporting individuals in performing daily activities and maintaining independence~\cite{Borg01032011}. Advances in artificial intelligence, machine learning, and sensing technologies have enabled the development of systems that can interact with users, monitor environments, and support decision-making~\cite{coeckelbergh2010health}. These technologies can be broadly categorized into virtual assistive systems, such as voice-enabled digital assistants, and robotic assistive systems that provide physical support in real-world environments~\cite{giachos2023inquiring}.

Virtual assistive systems depend on cloud-based services, natural language processing, and speech recognition to interpret user commands and execute tasks~\cite{Anand2025}. Such systems process large volumes of personal data. Thus, they raise concerns related to privacy, unauthorized access, and adversarial manipulation of voice inputs~\cite{Cheng2022}. In contrast, robotic assistive systems operate as cyber-physical systems that integrate perception, decision-making, and actuation~\cite{Omiyale2025}. While they offer greater autonomy and physical support, robotic assistive systems introduce additional risks, including sensor spoofing, adversarial perception, command injection, and threats to human safety.

Despite rapid progress in this area, existing literature has largely examined these systems in isolation and has not sufficiently addressed their comparative security challenges. Thus, a comparative perspective is essential, as virtual systems primarily face information security and privacy risks. At the same time, robotic systems must also address cyber-physical and safety concerns.

Unlike prior work that treats virtual assistants and robotic assistive systems separately, this paper develops a unified comparative framework for analyzing both, focusing on common security dimensions. This includes user interaction channel, data sensitivity, attacker access path, trust boundary, safety criticality, and mitigation priority. Such framing enables a better systematic comparison between digital and cyber-physical assistive systems and clarifies where their risks overlap and diverge. In addition, the developed framework explicitly incorporates trust boundaries, attacker capabilities, and cross-system risk characterization to support structured comparative reasoning.

In this paper, we present a comparative analysis of security and privacy challenges in virtual and robotic assistive systems. We examine their architectural differences, attack surfaces, and threat models, and discuss implications for system security and user safety. The main contributions are summarized as follows:
\begin{enumerate}
    \item A comparative framework for analyzing security and privacy challenges across virtual and robotic assistive technologies.
    \item A unified threat model that captures threat actors, attack surfaces, and protected assets across digital and cyber-physical assistive systems.
    \item A taxonomy of common attack types and their potential impacts on both classes of assistive systems.
    \item Security design recommendations for developing more resilient and trustworthy assistive technologies.
\end{enumerate}


\section{Related Work}

\subsection{Assistive Technologies and Intelligent Assistance}

Assistive systems have been significantly advanced in the last decade~\cite{paper1,Cowan_2012}. These systems are designed to enhance the quality of life. Current assistive systems range from software-based virtual assistants to physically embodied robotic systems operating in real-world environments.

Assistive technologies encompass a broad range of intelligent systems~\cite{de2022artificial}. This includes voice-based assistants, healthcare monitoring systems, assistive service robots, rehabilitation robots, and smart mobility aids. Each category serves distinct functions depending on user needs and operational settings. Table~\ref{tab:assistive_types} summarizes different assistive technology types and their primary tasks.

Virtual assistive systems are widely used in smart homes and healthcare environments~\cite{wang2025smart,amiri2025ai,giansanti2025integrating}. They rely on speech recognition, natural language processing, and cloud-based processing to interpret user commands and deliver personalized assistance~\cite{Kchaou2025}. Their ease of interaction makes them suitable for individuals with mobility or cognitive limitations. However, continuous audio monitoring and cloud connectivity introduce notable privacy and security concerns~\cite{adil2026internet,abba2024enabling}.

Assistive robotic systems represent a class of cyber-physical technologies that provide physical interaction and environmental support~\cite{mollaret2016multi,miller2006assistive}. These systems integrate sensing, computation, and actuation to assist with tasks such as object retrieval, navigation, and safety monitoring. In contrast, they are expanding the system attack surface due to their tight coupling with physical environments~\cite{ringwald2023should,Neupane2024,Belk2020,Zhang2019,Anniappa2021}.

\begin{table}[t]
\centering
\caption{Types of Assistive Technologies and Primary Tasks}
\label{tab:assistive_types}
\begin{tabular}{|p{2.6cm}|p{3.6cm}|p{5.6cm}|}
\hline
\textbf{Assistive Technology Type} & \textbf{Example Systems} & \textbf{Primary Tasks} \\
\hline
Virtual Assistants & Amazon Alexa, Google Assistant, Apple Siri & Voice interaction, smart home control, information retrieval, reminders, and daily task assistance \\
\hline
Health Monitoring Systems & Remote health monitoring systems, wearable devices & Monitoring physiological signals, tracking health metrics, and alerting caregivers or medical personnel \\
\hline
Service Robots & Domestic service robots, smart home robots & Object retrieval, mobility assistance, environmental monitoring, and task automation \\
\hline
Rehabilitation Robots & Exoskeletons, robotic therapy systems & Physical rehabilitation, movement assistance, and motor skill recovery \\
\hline
Socially Assistive Robots & Companion robots such as Paro or Pepper & Social interaction, cognitive stimulation, and emotional support \\
\hline
Smart Mobility Aids & Intelligent wheelchairs and navigation aids & Autonomous navigation, obstacle avoidance, and mobility support \\
\hline
\end{tabular}
\end{table}

\subsection{Security and Privacy Vulnerabilities in Virtual Assistive Systems}

Virtual assistive systems have attracted significant research attention due to their reliance on user data and cloud-based infrastructure~\cite{Modi2020}. Prior work shows that voice-based assistants are vulnerable to adversarial audio attacks, where crafted signals manipulate speech recognition systems into executing unintended commands~\cite{Yan2022}. These attacks may be embedded in background audio or ultrasonic signals that remain imperceptible to users.

Another major concern is unauthorized access to stored voice data and interaction logs~\cite{Tabbassum2024}. Because these systems continuously collect and transmit audio data to remote servers, they introduce substantial risks of data leakage and privacy violations. Attackers may exploit weaknesses in communication protocols, cloud storage, or authentication mechanisms to access sensitive information~\cite{Li2023}.

Replay attacks are also a common threat, where previously recorded voice commands are used to trigger system actions without user consent~\cite{monge2023defining}. Many systems lack robust speaker verification or contextual validation, making it difficult to distinguish legitimate commands from malicious ones~\cite{al2025comprehensive}. These vulnerabilities highlight the need for stronger authentication, encryption, and privacy-preserving mechanisms.

\subsection{Security Challenges in Assistive Robotic Systems}

Assistive robotic systems operate in physical environments and therefore face security risks beyond traditional information security concerns~\cite{Haskard2025}. These systems function as cyber-physical systems in which software components interact directly with sensors, control mechanisms, and physical actuators~\cite{bhardwaj2025securing}. Consequently, cyberattacks may lead not only to data breaches but also to unintended physical actions that compromise user safety.

Sensor spoofing is a critical vulnerability in robotic systems. By manipulating sensor inputs, adversaries can deceive perception algorithms and cause misinterpretation of objects, human actions, or environmental conditions~\cite{Shaik2025}. In assistive scenarios, such perception errors may lead to incorrect task execution, navigation failures, or unsafe interactions~\cite{Adesiji2025,Vasic2013,Guiochet2017}.

Command injection attacks target communication channels between software modules and actuators. Exploiting vulnerabilities in control interfaces or middleware, attackers may alter commands or inject malicious instructions, disrupting task execution or modifying system behavior.

Robotic systems are also exposed to network-based threats due to their reliance on wireless communication with cloud services and smart devices~\cite{Hu2012,GonzlezAlonso2012,Chen2018,ozer2020cloud,CastroAntonio2019,georgoulas2014home,SeungHoBaeg2007}. Weak authentication and insecure protocols may allow adversaries to intercept, modify, or inject data into control processes~\cite{Yaacoub2021,clark2017cybersecurity,GuerreroHigueras2018}.

Because assistive robots operate in close proximity to users, security breaches may result in both privacy violations and physical safety hazards~\cite{marchang2022assistive}. This dual impact distinguishes robotic systems from purely digital assistive systems.

Although both virtual and robotic assistive systems provide valuable support, their architectural differences lead to distinct security and privacy challenges. Virtual systems primarily face risks related to data privacy, unauthorized access, and adversarial input manipulation. In contrast, robotic systems introduce additional cyber-physical risks, including sensor spoofing, perception manipulation, and unsafe actuation. Table~\ref{tab:security_comparison} summarizes the key differences in their security limitations.

\begin{table}[t]
\centering
\caption{Security Limitations of Virtual and Robotic Assistive Systems}
\label{tab:security_comparison}
\begin{tabular}{|p{2.7cm}|p{4.5cm}|p{4.5cm}|}
\hline
\textbf{Category} & \textbf{Virtual Assistive Systems} & \textbf{Robotic Assistive Systems} \\
\hline
System Type & Digital and cloud-based systems & Cyber-physical systems interacting with physical environments \\
\hline
Primary Interaction Interface & Voice commands and natural language interaction & Sensors, cameras, and physical actuation \\
\hline
Data Collected & Voice recordings, user preferences, behavioral patterns & Video streams, environmental data, human activity information \\
\hline
Major Privacy Risks & Cloud data leakage, unauthorized data access, user profiling & Exposure of environmental monitoring data and user activities \\
\hline
Common Attack Types & Voice spoofing, replay attacks, adversarial audio attacks & Sensor spoofing, adversarial perception, command injection \\
\hline
Communication Vulnerabilities & Interception of cloud communications and API exploitation & Network-based attacks on robot communication channels \\
\hline
Authentication Challenges & Weak speaker verification and identity validation & Unauthorized device access and control channel exploitation \\
\hline
Safety Implications & Unauthorized execution of commands or device manipulation & Potential physical harm due to unsafe robot actions \\
\hline
Impact of Security Breach & Privacy violations and unauthorized control of smart devices & Physical safety risks and manipulation of robotic behavior \\
\hline
\end{tabular}
\end{table}


\section{Comparative Architecture of Virtual and Robotic Assistive Systems}

\begin{figure}[t]
\centering
\includegraphics[width=6cm, height = 8cm]{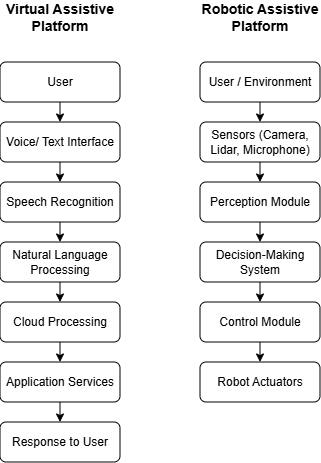}
\caption{The comparative architecture of virtual and robotic assistive systems and their attack surfaces.}
\label{fig:assistive_architecture}
\end{figure}

Virtual assistive systems typically follow a cloud-centric architecture in which user interactions are processed by speech recognition and natural language processing modules. Users mainly interact with the system via voice or text interfaces. Then the captured data is transmitted to cloud services, where machine learning models interpret the command and generate appropriate responses. The system architecture generally includes several fundamental components: an input interface, a speech recognition module, a natural language processing engine, a cloud-based processing infrastructure, and an application service layer that executes user requests. Figure \ref{fig:assistive_architecture} shows the reliance on cloud connectivity and centralized data processing creates potential vulnerabilities related to data interception, unauthorized access, and adversarial manipulation of input commands.

Assistive robotic systems depend on a distributed cyber-physical architecture that integrates sensing, perception, decision-making, and physical actuation~\cite{Singhal2023}. These systems collect environmental and user data through various sensors such as cameras, depth sensors, microphones, and tactile sensors. The captured data is processed by perception modules that interpret the surrounding environment and user activities. Based on the processed information, decision-making algorithms determine appropriate actions, which are then executed through robotic actuators responsible for navigation, manipulation, or interaction tasks. Figure \ref{fig:assistive_architecture} shows that integrating sensing and actuation components expands the system’s attack surface, as adversaries may target sensor inputs, perception algorithms, communication channels, or control systems.

The architectural differences between virtual and robotic assistive systems lead to different security and privacy challenges~\cite{brunete2021smart}. Virtual systems primarily expose vulnerabilities related to data privacy, cloud communication, and user authentication. However, robotic systems introduce further vulnerabilities in cyber-physical interactions. These vulnerabilities include sensor spoofing, perception manipulation, and unsafe actuation. Understanding these architectural differences is essential for identifying potential attack surfaces and developing effective security mechanisms for assistive technologies.


\section{Threat Model}\label{sec:modelThreat}

\begin{figure}[t]
	\centering
	\includegraphics[width=\textwidth]{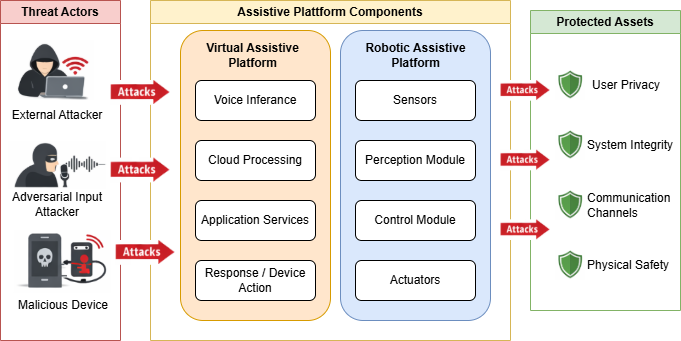}
	\caption{Threat model for virtual and robotic assistive systems.}
	\label{fig:threat_model}
\end{figure}

Assistive technologies operate through continuous interaction among users, devices, sensors, and cloud services~\cite{mulfari2015achieving,elsayed2022speech}. This creates multiple attack surfaces that may compromise user privacy, system integrity, and, in robotic systems, physical safety. To analyze these risks, we define a threat model that captures key assets, threat actors, attack surfaces, and attack goals across both virtual and robotic assistive systems~\cite{oruma2022systematic}. Figure~\ref{fig:threat_model} shows the threat model considered in this study.

\subsection{Assets and Security Objectives}

Assistive systems process sensitive data and, in some cases, interact directly with physical environments. The primary assets include user data, system functionality, communication channels, and physical safety~\cite{Giannetsos2011}. User data encompasses voice recordings, behavioral patterns, environmental observations, and interaction histories. System functionality must remain reliable to ensure correct interpretation of user commands and environmental inputs~\cite{Hamidi2018}. Communication channels are also critical, as both virtual and robotic systems rely on network connectivity to exchange data among sensors, processing modules, and cloud services. In robotic systems, physical safety constitutes an additional objective, as compromised perception or control may lead to unsafe actions~\cite{Omiyale2025}.

\subsection{Threat Actors}

Threat actors targeting assistive systems include external attackers, malicious insiders, compromised devices, and adversaries performing input manipulation attacks~\cite{abou2024cyber}. External attackers may exploit vulnerabilities in communication protocols, cloud services, or authentication mechanisms to gain unauthorized access~\cite{villalon2022taxonomy}. In smart environments, compromised devices may inject malicious commands or interfere with system communication. Input manipulation attacks directly target system interfaces, such as adversarial voice inputs in virtual assistants or manipulated sensor inputs in robotic systems~\cite{brunete2021smart}.

\subsection{System Assumptions and Trust Boundaries}

To refine the threat model, we define system assumptions and trust boundaries across assistive systems. Core components, including the local device operating system and authenticated user profiles, are assumed to operate within a trusted domain. However, several components reside in partially trusted or untrusted environments.

Trust boundaries arise at key interaction points, including: (i) the user-device interface, where adversarial inputs such as manipulated voice or sensor data may be introduced; (ii) the device-to-cloud communication channel, which may be subject to interception or tampering; (iii) the perception-control-actuation pipeline in robotic systems, where errors may propagate into unsafe physical actions; and (iv) third-party service integrations, where external APIs may introduce additional vulnerabilities.

External networks, surrounding environments, and connected IoT devices are considered untrusted and may serve as entry points for adversarial actions. These boundaries define where security controls must be enforced to preserve system integrity and safety.

\subsection{Attacker Capabilities and Constraints}

We consider multiple classes of attackers with varying capabilities. A nearby attacker may inject adversarial audio signals, manipulate visual inputs, or influence environmental conditions to affect perception. A network-based attacker may intercept, modify, replay, or inject communication traffic between system components. Further, a compromised-device attacker may exploit insecure IoT devices or smart home components to issue unauthorized commands.

We assume that attackers do not initially possess full system control but may exploit vulnerabilities to escalate their capabilities. In virtual assistive systems, attacker objectives typically involve triggering unauthorized commands or accessing sensitive data. In robotic assistive systems, these objectives extend to manipulating perception and control processes, potentially resulting in unsafe physical actions.

These capability assumptions enable a more precise evaluation of attack feasibility and impact across both system classes. The defined attacker models are used in Section~\ref{challenges} to support comparative analysis.

\subsection{Attack Surfaces}
The attack surface of assistive systems depends on their architecture. As shown in Figure~\ref{fig:assistive_architecture}, virtual systems expose attack surfaces associated with user interfaces, cloud services, APIs, and communication channels~\cite{chimuco2023secure}. Robotic systems present a broader attack surface due to the integration of sensing, perception, control, and actuation~\cite{krausz2019survey}. In particular, manipulation of sensor inputs may lead to incorrect perception, navigation errors, or unsafe interactions with users.

\subsection{Attack Goals}
Adversaries may aim to access sensitive user information, disrupt system functionality, or manipulate system behavior~\cite{li2020adversarial}. In virtual assistive systems, this typically results in privacy violations or unauthorized command execution. In robotic systems, attacks may additionally alter perception or control processes, introducing risks to both system integrity and physical safety.

\begin{table}[h]
\centering
\caption{Common Attacks and Potential Impacts in Assistive systems}
\label{tab:attack_taxonomy}
\begin{tabular}{|p{2cm}|p{2cm}|p{3.8cm}|p{3.8cm}|}
\hline
\textbf{Attack Type} & \textbf{Target System} & \textbf{Attack Description} & \textbf{Potential Impact} \\
\hline
Adversarial Voice Commands & Virtual Assistive Systems & Maliciously crafted audio signals designed to manipulate speech recognition systems & Unauthorized command execution and manipulation of connected smart devices \\
\hline
Replay Attacks & Virtual Assistive Systems & Previously recorded voice commands replayed to trigger system actions & Unauthorized system activation and user impersonation \\
\hline
Cloud Data Interception & Virtual Assistive Systems & Interception of communication between devices and cloud services & Leakage of sensitive user data and privacy violations \\
\hline
Sensor Spoofing & Robotic Assistive Systems & Manipulation of sensor inputs such as cameras or depth sensors & Incorrect environmental perception and unsafe robot behavior \\
\hline
Command Injection & Robotic Assistive Systems & Unauthorized modification of control commands transmitted to robotic actuators & Manipulation of robot actions and disruption of assistance tasks \\
\hline
Adversarial Perception Attacks & Robotic Assistive Systems & Adversarial inputs designed to deceive machine learning perception models & Misclassification of objects or human activities leading to incorrect system decisions\\
\hline
Network-Based Attacks & Both systems & Exploitation of vulnerabilities in wireless communication or connected devices & Data manipulation, unauthorized access, or system disruption \\
\hline
\end{tabular}
\end{table}

\section{Security and Privacy Challenges in Assistive Systems}\label{challenges}

Assistive systems introduce a range of security and privacy challenges due to their reliance on artificial intelligence, cloud connectivity, and sensor-based perception. These challenges differ across virtual and robotic systems due to their architectural and operational characteristics.

\subsection{Security Challenges in Virtual Assistive Systems}

Virtual assistive systems rely on voice-based interaction and cloud processing to interpret user commands and deliver services. While this architecture enables flexibility and scalability, it introduces vulnerabilities related to input manipulation, cloud communication, and data privacy~\cite{MarshMilne2024}.

Adversarial voice manipulation is a major threat in these systems~\cite{Zhang2019}. Malicious audio signals can be crafted to deceive speech recognition models into interpreting unintended commands~\cite{paperx}. Such inputs may be embedded in background audio or transmitted through ultrasonic signals that remain inaudible to users~\cite{Mao2020,Roy2017}. As a result, attackers may initiate unauthorized actions, including taking control of connected devices or accessing sensitive information.

Replay attacks represent another common vulnerability, where previously recorded commands are reused to activate system functions without user consent~\cite{Kamel_2025}. Because many systems rely on command recognition without strong speaker authentication, they remain susceptible to impersonation attacks~\cite{MarshMilne2024,MichelettoDeepfakeVC}.

Privacy risks are posing significant challenges~\cite{Grabler2024PrivacyBD}. These systems continuously collect and transmit user data to cloud services for processing and storage~\cite{Sharma2017}. Compromised communication channels or cloud infrastructure may expose sensitive information such as conversations, behavioral patterns, and household activity~\cite{Avenolu2022}. Weak encryption or inadequate data management policies further increase this risk.

Finally, vulnerabilities may arise from insecure application programming interfaces (APIs) used to integrate third-party services~\cite{Bolton2021}. Weak authentication or poorly designed interfaces may allow attackers to manipulate functionality or gain unauthorized access to user data.

Table~\ref{tab:attack_taxonomy} summarizes common attack types and their associated impacts.

\subsection{Security Challenges in Robotic Assistive systems}
The integration of sensing, perception, and actuation significantly expands the risk profile of assistive robotic systems. These systems operate as cyber-physical systems, where security breaches may produce both digital and physical consequences~\cite{thakur2025physical,khalid2018security,Omiyale2025}.

Under the threat model defined in Section~\ref{sec:modelThreat}, sensor spoofing is particularly critical because system behavior depends directly on environmental perception. Adversaries may manipulate visual or sensor inputs to influence perception results. This can cause a misclassification of objects or human activities~\cite{Bilika_2023}. In assistive contexts, such errors may directly affect safety-critical tasks, leading to unsafe behavior.

Command injection attacks represent another key vulnerability. By exploiting communication channels between software modules and actuators, attackers may alter control signals or disrupt task execution. In environments where robots operate in proximity to users, such manipulation can result in unintended physical actions and safety hazards~\cite{vasic2013safety}.

Robotic systems are also exposed to network-based threats due to their reliance on wireless communication with cloud services and smart devices~\cite{Bilika_2023}. Weak authentication, insecure protocols, or compromised devices may enable attackers to intercept or manipulate system data.

In addition, machine learning models used in perception and decision-making are susceptible to adversarial inputs~\cite{brendel2017decision,maqsood2023autonomous}. These attacks may degrade system performance and introduce risks to both task execution and user safety.

\subsection{Comparative Analysis of Security Risks}

Virtual and robotic assistive systems share concerns related to privacy, communication security, and unauthorized access, but their risk profiles differ substantially. Virtual systems are primarily exposed to threats involving data privacy, cloud communication, and authentication weaknesses~\cite{tabrizchi2020survey,sgandurra2016evolution}. In contrast, robotic systems introduce additional cyber-physical risks, including sensor manipulation, control compromise, and unsafe actuation.

Table~\ref{tab:security_comparison} summarizes these differences at a system level. However, beyond qualitative distinctions, the threat model enables a structured comparison of attack likelihood, impact, and safety implications.

To operationalize this comparison, we evaluate representative attack types across both systems using three criteria: likelihood of occurrence, impact on system functionality, and safety criticality. These criteria reflect both cybersecurity and cyber-physical risk dimensions.

\begin{table}[t]
\centering
\caption{Qualitative Risk Comparison Across Virtual and Robotic Assistive systems}
\label{tab:risk_comparison}
\begin{tabular}{|p{2.3cm}|p{1.8cm}|p{1.8cm}|p{1.8cm}|p{1.8cm}|p{2.2cm}|}
\hline
\textbf{Threat} & \textbf{Virtual Likelihood} & \textbf{Virtual Impact} & \textbf{Robotic Likelihood} & \textbf{Robotic Impact} & \textbf{Safety Criticality} \\
\hline
Adversarial Voice Commands & High & Medium & Low & Medium & Low \\
\hline
Replay Attacks & High & Medium & Low & Low & Low \\
\hline
Cloud Data Interception & Medium & High & Medium & High & Low \\
\hline
Sensor Spoofing & Low & Low & High & High & High \\
\hline
Command Injection & Low & Medium & Medium & High & High \\
\hline
Adversarial Perception & Low & Low & High & High & High \\
\hline
Network-Based Attacks & Medium & Medium & Medium & High & Medium \\
\hline
\end{tabular}
\end{table}

As shown in Table~\ref{tab:risk_comparison}, robotic assistive systems exhibit higher impact and safety criticality for perception and control-related attacks, whereas virtual systems are more susceptible to input manipulation and privacy-oriented threats. This contrast highlights the fundamentally different risk structures introduced by cyber-physical interaction and underscores the need for system-specific security strategies.

\section{Security Design Recommendations for Assistive Technologies}

The challenges discussed in Section~\ref{challenges} highlight the need for protection mechanisms that address both digital and cyber-physical risks. Because assistive technologies process sensitive data and often interact closely with users, their design must protect confidentiality, integrity, availability, and, in robotic systems, physical safety. This section outlines key recommendations for improving the resilience of assistive systems.

\subsection{Secure Input Validation and Authentication}
User input interfaces are a major attack surface, particularly in voice-based virtual assistants. Adversarial voice manipulation and replay attacks show that systems relying only on command recognition are vulnerable to unauthorized actions. To mitigate these risks, assistive systems should implement stronger authentication mechanisms, such as speaker verification, contextual checks, and multi-factor authentication, for sensitive commands. In addition, anomaly detection techniques can help identify suspicious inputs that deviate from normal usage patterns.

\subsection{Privacy-Preserving Data Processing}
Many virtual assistive systems rely on cloud-based processing, increasing exposure to sensitive user data. Privacy risks can be reduced through on-device processing, data minimization, and encrypted communication. Edge computing can further limit unnecessary data transmission by performing selected tasks locally, thereby reducing opportunities for interception or misuse.

\subsection{Robust Perception and Sensor Security}
In robotic assistive systems, perception depends heavily on sensor inputs. To reduce the risk of manipulated or misleading inputs, assistive robots should employ sensor redundancy, cross-validation, and anomaly detection. Model robustness can also be improved through adversarial training and verification techniques for perception algorithms.

\subsection{Secure Communication and System Integration}
Assistive systems often operate in interconnected smart environments, making communication security essential. Protection mechanisms should include end-to-end encryption, secure device pairing, access control, and continuous monitoring of network activity to reduce the risk of unauthorized access or data manipulation.

\subsection{Human-in-the-Loop Safety Mechanisms}
As assistive technologies interact directly with users, safety must remain a central design goal. In robotic systems, human-in-the-loop mechanisms enable users or caregivers to override abnormal behavior. Fail-safe responses can further reduce the impact of compromised perception or control.

\section{Discussion}

Our conducted analysis in this paper demonstrates that assistive technologies require a multi-layered security perspective that accounts for both digital and cyber-physical risks. While prior work has typically examined virtual assistants and robotic systems independently, this paper provides a unified comparative framework that emphasizes how differences in interaction modality, system architecture, and physical embodiment influence security risk.

A critical insight from our comparative analysis is that risk in assistive technologies is not only a function of vulnerability but also of system context. Virtual assistive systems are primarily affected by privacy, authentication, and cloud-related threats, while robotic assistive systems present additional safety-critical risks due to their interaction with the physical environment. This distinction becomes particularly crucial when evaluating attack impact, as cyber-physical systems may translate digital compromise into physical consequences.

The qualitative risk assessment also shows that similar attack classes can have fundamentally different implications across systems. For example, input manipulation in virtual systems primarily affects privacy and device control. In contrast, perception manipulation in robotic systems may directly impact user safety. These findings reinforce the need for system-aware security design strategies rather than universal solutions.

Another important consideration is the balance between security and usability. Assistive technologies are often designed for users with limited mobility or technical expertise. As a result, security mechanisms must provide strong protection without introducing barriers to accessibility. Achieving this balance remains a critical challenge for real-world deployment.

Further, our paper highlights the need for more formal and quantitative approaches to evaluating security risks in assistive technologies. While this study provides a structured qualitative framework, future work can extend this approach through empirical validation, simulation-based analysis, or dataset-driven evaluation of attack scenarios.

\section{Conclusion}

This paper presented a comparative analysis of security and privacy challenges in virtual and robotic assistive systems. Unlike prior studies that treat these systems independently, this work introduced a unified framework for analyzing assistive technologies across shared security dimensions, including interaction modality, data sensitivity, attack surfaces, trust boundaries, and safety implications.

By integrating architectural analysis with a structured threat model, our paper identified key differences in how vulnerabilities manifest across digital and cyber-physical systems. The results show that virtual systems are primarily exposed to privacy and authentication risks. On the other hand, robotic systems introduce additional safety-critical threats due to their reliance on sensors and physical actuation.

To support this comparison, the paper proposed a qualitative risk assessment that evaluates attack likelihood, impact, and safety criticality across both system types. This analysis provides a systematic basis for understanding how similar attack vectors can produce fundamentally different consequences depending on system context.

The paper also outlined design recommendations to improve the resilience of assistive technologies, including stronger authentication mechanisms, privacy-preserving data processing, robust perception systems, secure communication protocols, and human-in-the-loop safety controls.

As assistive technologies continue to evolve, ensuring their security requires approaches that jointly consider digital vulnerabilities and physical safety risks. Future work may extend this framework through empirical validation, formal risk modeling, and the development of standardized security evaluation methodologies for assistive systems.

\bibliographystyle{splncs04}
\bibliography{references}

\end{document}